\documentclass[aps,prl,reprint,showpacs,superscriptaddress]{revtex4-1}
\usepackage{graphicx}
\usepackage{dcolumn}
\usepackage{bm}
\usepackage{trfsigns}
\usepackage{amsmath}
\usepackage{comment}
\usepackage{dsfont}
\usepackage{enumerate}
\usepackage{amssymb}
\usepackage{braket}
\usepackage{enumitem}
\usepackage{nicefrac}
\usepackage{dsfont}
\usepackage{mathtools}
\usepackage{color}

\begin{document}


\title{Experimental demonstration of direct path state characterization by strongly measuring weak values in a matter-wave interferometer}


\author{Tobias Denkmayr}
\email[]{tdenkmayr@ati.ac.at}
\affiliation{Atominstitut, TU Wien, Stadionallee 2, 1020 Vienna, Austria}
\author{Hermann Geppert}
\affiliation{Atominstitut, TU Wien, Stadionallee 2, 1020 Vienna, Austria}
\author{Hartmut Lemmel}
\affiliation{Atominstitut, TU Wien, Stadionallee 2, 1020 Vienna, Austria}
\affiliation{Institut Laue-Langevin, 6, Rue Jules Horowitz, 38042 Grenoble Cedex 9, France}
\author{Mordecai Waegell}
\affiliation{Institute for Quantum Studies, Chapman University, Orange, CA 92866, USA}
\author{Justin Dressel}
\affiliation{Institute for Quantum Studies, Chapman University, Orange, CA 92866, USA}
\affiliation{Schmid College of Science and Technology, Chapman University, Orange, CA 92866, USA}
\author{Yuji Hasegawa}
\affiliation{Atominstitut, TU Wien, Stadionallee 2, 1020 Vienna, Austria}
\author{Stephan Sponar}
\affiliation{Atominstitut, TU Wien, Stadionallee 2, 1020 Vienna, Austria}


\date{\today}

\begin{abstract}
A novel method was recently proposed and experimentally realized for characterizing a quantum state by directly measuring its complex probability amplitudes in a particular basis using so-called weak values. Recently Vallone and Dequal showed theoretically that weak measurements are not a necessary condition to determine the weak value [Phys.~Rev.~Lett.~{\bf116,} 040502 (2016)]. Here we report a measurement scheme used in a matter-wave interferometric experiment in which the neutron path system's quantum state was characterized via direct measurements using both strong and weak interactions. Experimental evidence is given that strong interactions outperform weak ones. Our results are not limited to neutron interferometry, but can be used in a wide range of quantum systems.

\end{abstract}

\maketitle

Ever since Aharonov, Albert and Vaidman (AAV) introduced the weak value as ``\textit{a new kind of value for a quantum variable}''~\cite{aav88}, it has been a topic of intense debate~\cite{duck89,legget89}. While theoretical discussions about its physical interpretation still last today~\cite{svensson2013,ferrie2014,sokolovski2015,pusey2014,dressel2015}, the weak value is unequivocally a powerful experimental tool~\cite{dressel2014}: It can be used for high precision metrology by amplifying detector signals~\cite{hosten08} and as a conditioned average of generalized observable eigenvalues~\cite{johansen2004}, it offers new insights into quantum paradoxes, such as Hardy's paradox~\cite{hardy1992,lundeen09,yokota2009}, the three-box paradox~\cite{aharonov1991,resch2004} and the quantum Cheshire cat~\cite{aharonov13,denkmayr2014}. One can also take a pragmatic approach and simply treat the weak value as a complex number that is accessible by experiment, as done in direct state characterization~\cite{lundeen2011,salvail2013} to determine complex quantum state probability amplitudes in a particular basis.\\
The weak value of observable $\hat{\text{A}}$ of a quantum system is given by
\begin{align}
\braket{\hat{\text{A}}}_\mathrm{w}=\frac{\braket{\psi_\mathrm{f}|\hat{\text{A}}|\psi_\mathrm{i}}}{\braket{\psi_\mathrm{f}|\psi_\mathrm{i}}}\label{eqn:wv_define},
\end{align}
where $\Ket{\psi_\mathrm{i}}$ and $\ket{\psi_\mathrm{f}}$ are the initial (preselected) and final (postselected) system states respectively. To determine $\braket{\hat{\text{A}}}_\mathrm{w}$ a probe system, which serves as a measurement apparatus, has to be coupled to the observed system, leading to an entanglement between them. In the usual weak measurement approach only minimally disturbing interactions, between the quantum system and the measurement apparatus are regarded. However, as was recently pointed out theoretically~\cite{vallone2016,zhang2016}, the weakness of the interaction is not a necessary condition to obtain the weak value. Furthermore it was shown that strong measurements give a better direct measurement of the quantum wave function using the weak value.\\
Originally AAV constructed the weak value formalism in a non-relativistic quantum framework and hence it should be first and foremost applicable to massive quantum systems. Consequently they proposed a modified Stern-Gerlach experiment with spin-$\nicefrac{1}{2}$ particles to test their measurement scheme~\cite{aav88}. In general optical experiments with matter-waves provide excellent conditions to demonstrate the peculiarities of quantum mechanics \cite{rauch00,cronin2009,arndt2014}. However, due to the small coherence volume of massive particle beams, an experimental demonstration of a weak value's measurement in a simple massive-particle system proved to be difficult: the first experimental determination of a weak value was realized in a purely optical setup~\cite{ritchie91}. Significant improvements in the technique of neutron interferometry~\cite{geppert2014} made it possible to fully determine the weak value of a neutron's spin operator with high precision~\cite{sponar2015,sponar2016}. Neutron interferometry has been established as a powerful experimental method to investigate the foundations of quantum mechanics~\cite{rauch02,hasegawa03,pushin2011,hasegawa11,klepp2014,clark2015}. In combination with the novel weak value measurement scheme it offers an experimental window into previously inaccessible parts of massive quantum systems.\\
Here we present an experiment in which the neutron's path degree of freedom's (DOF)~\cite{basu2001} state vector is characterized using weak values. The observable of interest is the Pauli operator $\hat{\sigma}_z^\mathrm{p}$ and the neutron's spin DOF serves as a meter system. The measurement of $\braket{\hat{\sigma}_z^\mathrm{p}}_\mathrm{w}$ makes it possible to directly characterize the preselected path state. The weak values are obtained through weak and strong interactions and the precision as well as the accuracy of both experimental approaches are quantified. The experimental results support the statements made in~\cite{vallone2016}, that strong measurements indeed outperform weak ones.\\

The measurement scheme starts with the initial state
\begin{align}
\Ket{\Psi_\mathrm{i}}=\Ket{\mathrm{P}_\mathrm{i}}\Ket{\mathrm{S}_\mathrm{i}}&=\left(c_+\Ket{\mathrm{P}_z;+}+ c_-\Ket{\mathrm{P}_z;-}\right)\Ket{\mathrm{S}_x;+},\label{eqn:path_initial_state_ps}
\end{align}
where $\Ket{\mathrm{P}_\mathrm{i}}$ is the initial path and $\Ket{\mathrm{S}_\mathrm{i}}$ the initial spin state. $\Ket{\mathrm{P}_z;+}$ and $\Ket{\mathrm{P}_z;-}$ are the eigenstates of path $I$ and $II$ respectively, with the corresponding probability amplitudes $c_+$ and $c_-$. $\Ket{\mathrm{S}_x;+}$ denotes a spin state that is aligned along the positive $x$-axis. A general form for a preselected path state is given by
\begin{align}
{\Ket{\mathrm{P}_\mathrm{i}}=\cos\left(\frac{\theta}{2}\right)\Ket{\mathrm{P}_z;+}+\exp\left(\mathrm{i}\phi\right)\sin\left(\frac{\theta}{2}\right)\Ket{\mathrm{P}_z;-}},\label{eqn:preselected_general}
\end{align}
where $\phi$ represents the relative phase and $\theta$ the weight of the two eigenstates. The probability amplitudes $c_+$ and $c_-$ are linked to the general state by ${\phi=\operatorname{arg}\left(c_+\right)-\operatorname{arg}\left(c_-\right)}$ and ${\cos\theta=\left|c_+\right|^2-\left|c_-\right|^2}$.\\
Equation~(\ref{eqn:path_initial_state_ps}) describes a completely separable state. There is no coupling between the spin and path DOF. As a next step a coupling is created by a unitary evolution consisting of path dependent spin rotations. More precisely the spin is rotated by a certain angle $\alpha$ around the $z$-axis in the $xy$-plane with positive (clockwise) rotations in path $I$ and negative (counter clockwise) ones in path $II$. The interaction Hamiltonian for such a measurement is
\begin{align}
\hat{H}_\mathrm{int}&=-\vec{\mu}\cdot\vec{\mathrm{B}}~\hat{\Pi}_{z+}^\mathrm{p}+\vec{\mu}\cdot\vec{\mathrm{B}}~\hat{\Pi}_{z-}^\mathrm{p}\label{eqn:spin_hamiltonian}
\end{align}
where $\hat{\Pi}_{z\pm}^\mathrm{p}$ are the projection operators on the path eigenstates $\Ket{\mathrm{P}_z;+}$ and $\Ket{\mathrm{P}_z;-}$, $\vec{\mu}$ is the neutron's magnetic moment and $\vec{\mathrm{B}}=\left(0,0,B_z \right)$ an applied magnetic field.\\
The action of $\hat{H}_\mathrm{int}$ on the composite system $\ket{\Psi_\mathrm{i}}$ is described by an evolution operator 
\begin{align}
\ket{\Psi'}=\mathrm{e}^{\nicefrac{-\mathrm{i}}{\hbar}\int\hat{H}_\mathrm{int} \mathrm{d}t}\Ket{\Psi_\mathrm{i}}=\mathrm{e}^{\nicefrac{-\mathrm{i}\alpha\hat{\sigma}_z^\mathrm{s}\hat{\sigma}_z^\mathrm{p}}{2}}\Ket{\Psi_\mathrm{i}}.
\end{align}
The angle of rotation $\alpha$ is given by $\nicefrac{-2\mu\mathrm{B}_z\tau}{\hbar}$, where $\tau$ is the neutron's transit time in the magnetic field region. $\alpha$ is the relevant parameter for the interaction strength of the measurement. $\hat{\sigma}_z^\mathrm{s}$ is the generator of spin rotations around the $z$-axis. The Pauli operator $\hat{\sigma}_z^\mathrm{p}$ is given by $\hat{\sigma}_z^\mathrm{p}=\Ket{\mathrm{P}_z;+}\Bra{\mathrm{P}_z;+}-\Ket{\mathrm{P}_z;-}\Bra{\mathrm{P}_z;-}$.\\
In the standard weak measurement procedure~\cite{sponar2015} the evolution operator $\exp\left(\nicefrac{-\mathrm{i}\alpha\hat{\sigma}_z^\mathrm{s}\hat{\sigma}_z^\mathrm{p}}{2}\right)$ is series expanded around $\alpha$ and by neglecting higher orders of $\alpha$ an approximation for $\alpha\ll1$ is made. Here, however, the analytical relation
$\exp\left(\nicefrac{-\mathrm{i}\alpha\hat{\sigma}_z^\mathrm{s}\hat{\sigma}_z^\mathrm{p}}{2}\right)=\cos\left(\nicefrac{\alpha}{2}\right)-\mathrm{i}\hat{\sigma}_z^\mathrm{s}\hat{\sigma}_z^\mathrm{p}\sin\left(\nicefrac{\alpha}{2}\right)$
is used~\cite{vallone2016}. No approximation is made. Therefore the calculations hold for arbitrary interaction strengths. The analytic form of the state after the interaction is given by
\begin{align}
\Ket{\Psi'}&=\cos\left(\frac{\alpha}{2}\right)\Ket{\mathrm{P}_\mathrm{i}}\Ket{\mathrm{S}_x;+}-\mathrm{i}\hat{\sigma}_z^\mathrm{p}\sin\left(\frac{\alpha}{2}\right)\Ket{\mathrm{P}_\mathrm{i}}\Ket{\mathrm{S}_x;-}
\end{align}
The final step of the measurement scheme is the postselection. The path is postselected on the final state
\begin{align}
\ket{\mathrm{P}_\mathrm{f}}=\ket{\mathrm{P}_x;+}=\frac{1}{\sqrt2}\left(\ket{\mathrm{P}_z;+}+\ket{\mathrm{P}_z;-} \right).\label{eqn:post_path_state}
\end{align}
The action of the path postselection is equivalent to a projection onto $\ket{\mathrm{P}_\mathrm{f}}\bra{\mathrm{P}_\mathrm{f}}$. It leads to final state $\ket{\Psi_\mathrm{f}}$, which has the form 
\begin{align}
\Ket{\Psi_\mathrm{f}}&=\Braket{\mathrm{P}_\mathrm{f}|\mathrm{P}_\mathrm{i}}\left\lbrack \cos\left(\frac{\alpha}{2} \right)\Ket{\mathrm{S}_x;+}-\right.\nonumber\\
&\hspace{22mm}\left.-\mathrm{i}\sin\left(\frac{\alpha}{2}\right)\Braket{\hat{\sigma}_z^\mathrm{p}}_\mathrm{w}\Ket{\mathrm{S}_x;-}\right\rbrack\ket{\mathrm{P}_\mathrm{f}}.
\end{align}
Finally the weak value of $\hat{\sigma}_z^\mathrm{p}$ is determined by evaluating the pointer system. Projective measurements along the six spin directions $\pm x$, $\pm y$ and $\pm z$, yield six intensities ${\text{I}_{j\pm}=\left|\braket{\mathrm{S}_j;\pm|\Psi_\mathrm{f}} \right|^2}$ with $\left(j=x,y,z \right)$, which allow us to extract the imaginary and real part as well as the modulus of the path operator's weak value. It is straightforward to derive the relations
\begin{subequations}
\begin{alignat}{2}
&\operatorname{Re}\left(\Braket{\hat{\sigma}_z^\mathrm{p}}_\mathrm{w}\right)&&=\frac{1}{2}\cot\left(\frac{\alpha}{2}\right)\frac{\text{I}_{y+}-\text{I}_{y-}}{\text{I}_{x+}}\label{eqn:path_extract_re_allorder}\\
&\operatorname{Im}\left(\Braket{\hat{\sigma}_z^\mathrm{p}}_\mathrm{w}\right)&&=\frac{1}{2}\cot\left(\frac{\alpha}{2}\right)\frac{\text{I}_{z+}-\text{I}_{z-}}{\text{I}_{x+}}\label{eqn:path_extract_im_allorder}\\
&\left|\Braket{\hat{\sigma}_z^\mathrm{p}}_\mathrm{w}\right|&&=\cot\left(\frac{\alpha}{2}\right)\sqrt{\frac{\text{I}_{x-}}{\text{I}_{x+}}}\label{eqn:path_extract_abs_allorder}
\end{alignat}
\end{subequations}
which connect the intensities ${\text{I}_{j\pm}}$ to all components of $\braket{\hat{\sigma}_z^\mathrm{p}}_\mathrm{w}$. Due to the the choice of the meter's initial direction and the rotational axis for the spin path coupling, anisotropy emerges and $\text{I}_{x+}$ appears as a normalization factor. It has to be stressed that no approximations are made to derive this result: relations (\ref{eqn:path_extract_re_allorder}) to (\ref{eqn:path_extract_abs_allorder}) hold for any value of $\alpha$, i.e., for arbitrary measurement strengths. The above argument is not limited to the neutron's spin and path DOF, but it can be applied to any coupling between two two-level quantum systems.\\
Since $\braket{\hat{\Pi}_{z\pm}^\mathrm{p}}_\mathrm{w}=\nicefrac{\mathds{1}\pm\braket{\hat{\sigma}_z^\mathrm{p}}_\mathrm{w}}{2}$, the complete determination of the weak value of the Pauli operator $\hat{\sigma}_z^\mathrm{p}$ also gives the weak values of the projection operators on each path eigenstate. These in turn characterize the measured preselected path state~\cite{salvail2013}
\begin{align}
\Ket{\mathrm{P}_\mathrm{i}^\mathrm{m}}=\frac{\Braket{\hat{\Pi}_{z+}^\mathrm{p}}_\mathrm{w}\Ket{\mathrm{P}_z;+}+\Braket{\hat{\Pi}_{z-}^\mathrm{p}}_\mathrm{w}\Ket{\mathrm{P}_z;-}}{\sqrt{\left|\Braket{\hat{\Pi}_{z+}^\mathrm{p}}_\mathrm{w}\right|^2+\left|\Braket{\hat{\Pi}_{z-}^\mathrm{p}}_\mathrm{w}\right|^2}}\label{eqn:tomography_normalized}.
\end{align}
By denoting the normalization factor as ${\nu\equiv\nicefrac{1}{\sqrt{\left|\Braket{\hat{\Pi}_{z+}^\mathrm{p}}_\mathrm{w}\right|^2+\left|\Braket{\hat{\Pi}_{z-}^\mathrm{p}}_\mathrm{w}\right|^2}}}$, the probability amplitudes of $\ket{\mathrm{P}_\mathrm{i}^\mathrm{m}}$ are given by $c_+=\nu\braket{\hat{\Pi}_{z+}^\mathrm{p}}_\mathrm{w}$ and $c_-=\nu\braket{\hat{\Pi}_{z-}^\mathrm{p}}_\mathrm{w}$. They are directly proportional to quantities that are determined experimentally, namely to the path projection operators' weak values.
Using the weak value's definition given by Eq.~(\ref{eqn:wv_define}), as well as the pre and postselected path states of Eqs.~(\ref{eqn:preselected_general}) and (\ref{eqn:post_path_state}), respectively, one expects $\braket{\hat{\Pi}_{z\pm}^\mathrm{p}}_\mathrm{w}$ to be
\begin{align}
\braket{\hat{\Pi}_{z\pm}^\mathrm{p}}_\mathrm{w}=\frac{\braket{\mathrm{P}_\mathrm{f}|\hat{\Pi}_{z\pm}^\mathrm{p}|\mathrm{P}_\mathrm{i}}}{\braket{\mathrm{P}_\mathrm{f}|\mathrm{P}_\mathrm{i}}}=\frac{1}{2}\mp\frac{\mathrm{i}}{2}\tan\left(\frac{\phi}{2} \right),
\end{align}
if one assumes that $\theta=\nicefrac{\pi}{2}$ in Eq.~(\ref{eqn:preselected_general}), as is the case for a 50:50 beam splitter of a Mach-Zehnder type interferometer.\\

\begin{figure}[h!]
\includegraphics[width=1.0\columnwidth]{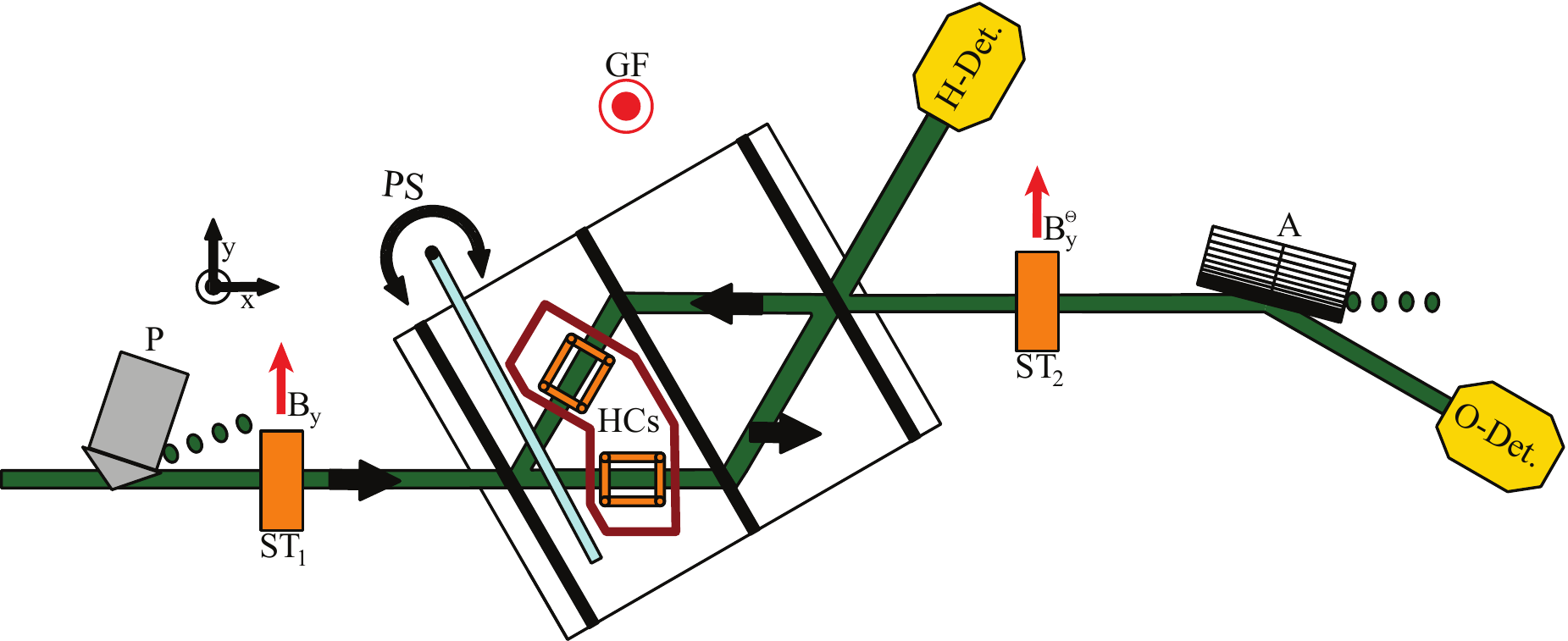}
\caption{Schematic drawing of the experimental setup (color online): The neutron beam passes magnetic prisms (P). To prevent depolarization a magnetic guide field (GF) is applied around the whole setup. Before the neutrons enter the interferometer a spin turner (ST$_1$) generates the initial spin state $\ket{\mathrm{S}_\mathrm{ i}}$. At the interferometer's first plate the neutron beam gets separated into path $I$ and $II$. In each beam path small coils in Helmholtz configuration (HCs) allow the manipulation of the neutron spin in the $xy$-plane. The phase shifter (PS) tunes the relative phase $\chi$ between path $I$ and $II$. The spin postselection is performed using a spin turner (ST$_2$) in combination with a CoTi supermirror (A). Two interfering beams leave the interferometer and only that in the forward direction is spin analyzed and detected by the O-detector (O-Det).\label{fig:setup}}
\end{figure}

\begin{figure}[ht!]
\includegraphics[width=1.0\columnwidth]{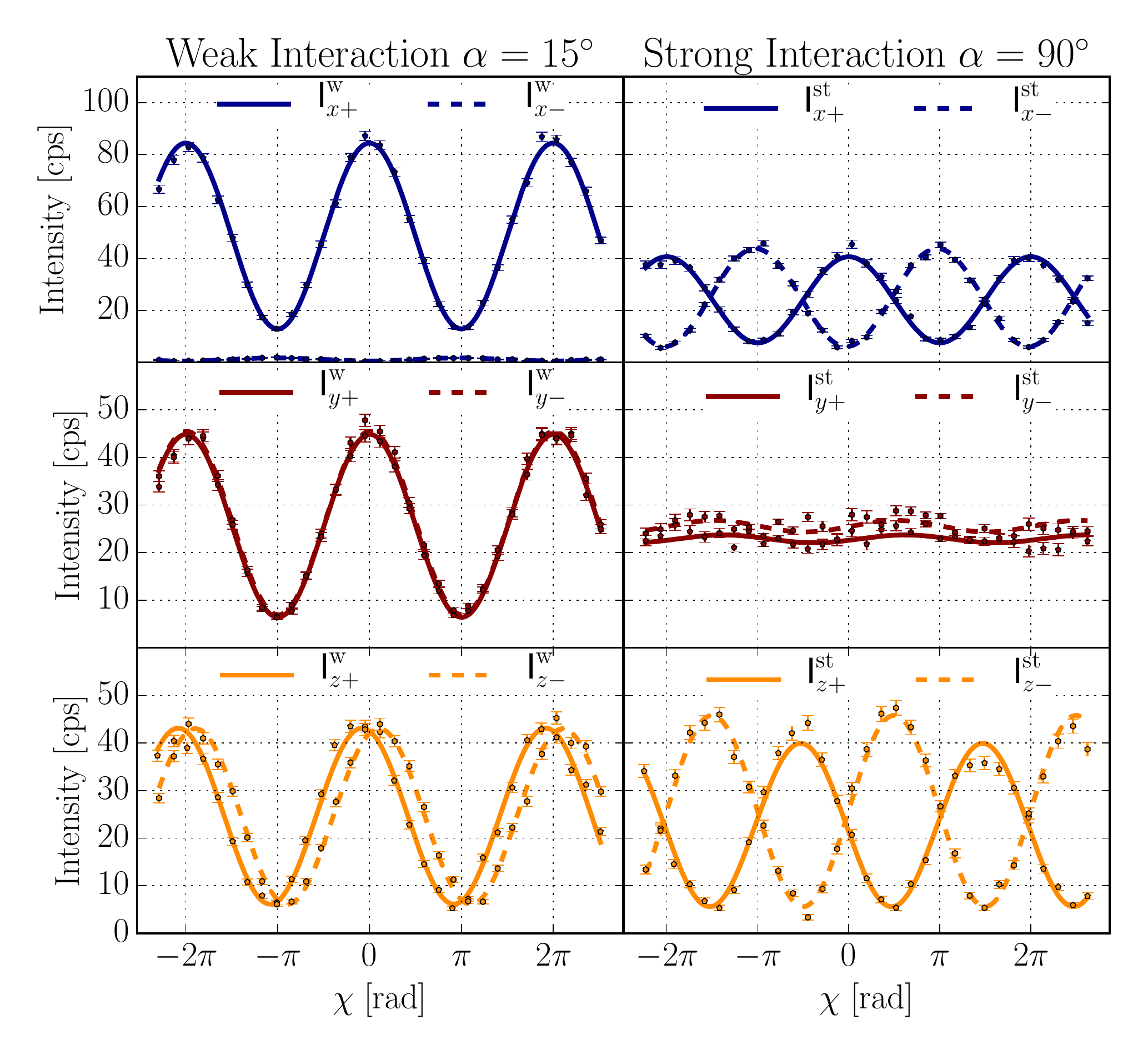}
\caption{Interference fringes of the weak measurement ($\alpha=15^\circ$) on the left side and those of the strong measurement ($\alpha=90^\circ$) on the right side (color online). The directions of spin analysis $\pm x$, $\pm y$ and $\pm z$ are shown in the top, middle and bottom row respectively. The solid and dashed lines show the least square fits for the + and - analysis directions respectively. The error bars show one standard deviation. The background has already been subtracted from all interferograms.\label{fig:interferograms}}
\end{figure}

A neutron interferometric experiment was performed at the beamline S18 at the high flux research reactor at the Institut Laue-Langevin (ILL) in Grenoble, France. A schematic drawing of the interferometric setup is depicted in FIG.~\ref{fig:setup}.\\
From a white neutron beam particles with a wavelength $\lambda_0=1.91$~\AA $(\lambda/ \lambda_0\sim 0.02)$ are selected by a triple bounce silicon perfect crystal monochromator and subsequently pass magnetic prisms. They deflect the spin down component out of the Bragg condition of the interferometer crystal, such that only spin up neutrons are accepted by the interferometer~\cite{kroupa2000}. The prisms polarize the neutron beam along the positive $z$-axis. For our experiment the degree of polarization was determined to be over 99\%. A DC coil turns the neutron spin by $\nicefrac{\pi}{2}$, due to Larmor precession within the coil, so that the spin is aligned along the positive $x$-axis. To tune the relative phase $\chi$, a parallel sided sapphire slab is inserted between the first and the second plate of the interferometer as a phase shifter.\\
After the phase shifter, the initial state $\ket{\Psi_\mathrm{i}}$ is generated and the preselection procedure is complete. Inside the interferometer a coil in Helmholtz configuration in each beam path enables us to perform path-dependent spin rotations, coupling the path and spin DOF~\cite{geppert2014}. The coils produce additional magnetic fields in the $\pm z$-direction that cause the neutron spins' Larmor precession frequency $\omega_\mathrm{L}$ to decrease or increase depending on the sign of the field. The strength of the magnetic field determines the magnitude of the rotation angle $\alpha$. The experiment is performed with two different values of $\alpha$. To test the interaction in a weak regime $\alpha$ is set to $15^\circ$. For the strong interactions $\alpha$ is set to $90^\circ$, which corresponds to the maximum measurement strength, due to the orthogonality of the spin states after the interaction.\\
At the interferometer's third plate the beams are recombined. By recombining the beams the path postselection is carried out. Only neutrons leaving the interferometer in the forward direction with a relative phase $\chi=0$ are spin analyzed. The spin analysis is performed by a second DC coil mounted on a translation stage in combination with a CoTi supermirror. Inside the coil a magnetic field $B_y^\Theta$ rotates the spin by a polar angle $\Theta$. Depending on the coil's position along the neutrons' trajectory, the azimuth angle $\Phi$ is tuned due to the spin's Larmor precession within the guide field. By applying different magnetic fields $B_y^\Theta$ inside the coil any polar angle $\Theta$ can be tuned. By placing the coil at different positions any azimuth angle $\Phi$ can be selected. Subsequently the supermirror array filters out all neutrons but those with a spin aligned along the selected $(\Phi,\Theta)$-direction. Finally the neutrons are detected by a $^3$He~counter (O-detector).\\
The intensity modulations $\text{I}_{j\pm}$ of both experimental runs are depicted in FIG.~\ref{fig:interferograms}.
The three left panels show the interferograms of the weak interaction ($\alpha=15^\circ$). For $\text{I}_{x+}^{\rm w}$ both the pre- and postselected spin state are $\ket{\mathrm{S}_x;+}$, leading to a large count rate. In contrast the count rate of $\text{I}_{x-}^{\rm w}$ is very low, due to the orthogonality of initial and final spin states. $\text{I}_{y\pm}^{\rm w}$ are identical and have in average half of the maximal count rate. $\text{I}_{z\pm}^{\rm w}$ are phase shifted to each other by two times $\alpha$ and they show the same average count rate. In the three right panels the interferograms for the strong interaction are shown. Due to the large spin rotation of $\alpha=\pm90^\circ$ in each beam path, $\text{I}_{x\pm}^{\rm st}$ now show the same average count rate, while being phase shifted by $\pi$. $\text{I}_{y\pm}^{\rm st}$ show only negligibly little contrast. The phase shift between $\text{I}_{z+}^{\rm st}$ and $\text{I}_{z-}^{\rm st}$ is now also expected to be $\pi$ and easy to resolve.

Some advantages of the strong measurement approach can already be seen in the differences between the interference fringes of the weak and the strong interaction measurement. To extract the imaginary part of $\braket{\hat{\sigma}_z^\mathrm{p}}_\mathrm{w}$ the intensities $\text{I}_{z\pm}$ and $\text{I}_{x+}$ are used. While $\text{I}_{x+}$ acts as a normalization factor, resolving the phase shift between $\text{I}_{z+}$ and $\text{I}_{z-}$ is crucial to determine $\operatorname{Im}\left(\braket{\hat{\sigma}_z^\mathrm{p}}_\mathrm{w}\right)$. Since this phase shift is expected to be two times $\alpha$, it is much harder to resolve in the weak interaction case. Similarly $\operatorname{Re}\left(\braket{\hat{\sigma}_z^\mathrm{p}}_\mathrm{w}\right)$ is extracted from the intensities $\text{I}_{y\pm}$ and $\text{I}_{x+}$. Because the weak value's real part is expected to be zero, $\text{I}_{y+}$ and $\text{I}_{y-}$ are very close to equal~\cite{sponar2015}. Furthermore $\text{I}_{y\pm}$ lose contrast for larger $\alpha$, due to the spin rotation inside the interferometer leading to an orthogonal spin state. For $\alpha=90^\circ$ the spin state is completely orthogonal and shows no contrast.\\
Finally the modulus of $\braket{\hat{\sigma}_z^\mathrm{p}}_\mathrm{w}$ is directly obtained from the $\text{I}_{x\pm}$ data. It is proportional to the square root of $\nicefrac{\text{I}_{x-}}{\text{I}_{x+}}$. The advantages of the strong interaction approach are intuitively understood. Here, the discrimination of the relevant signal from the background is crucial: For $\alpha=0^\circ$, $\text{I}_{x-}$ is also expected to be zero and the signal becomes larger with increasing $\alpha$. If $\alpha$ is kept small as in the weak measurement approach, it is very hard to discriminate the intensity from the background.

\begin{figure}[h!]
\includegraphics[width=1.0\columnwidth]{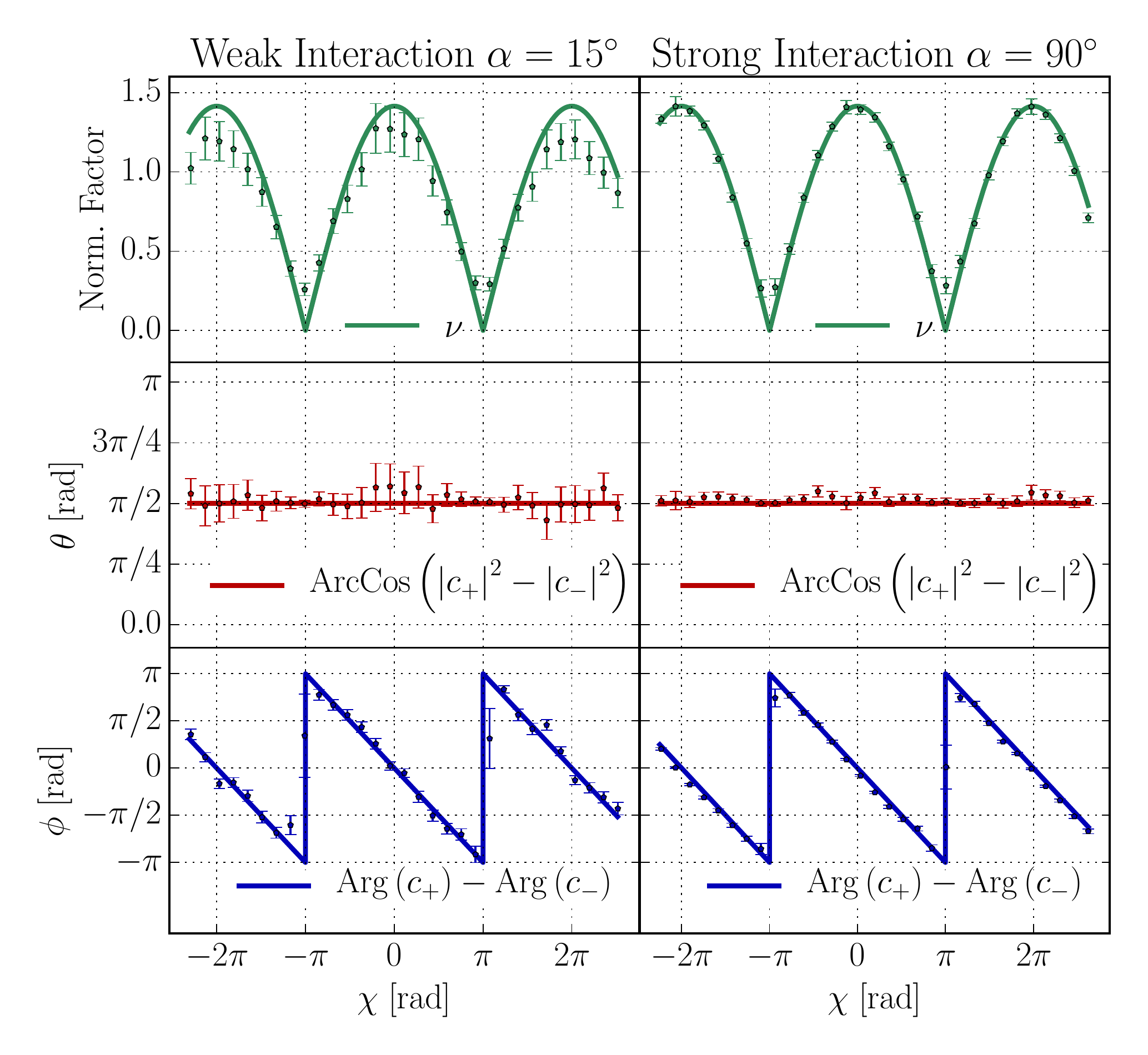}
\caption{Measurement results for the path state vector for both, the weak ($\alpha=15^\circ$) and the strong measurement ($\alpha=90^\circ$) case (color online): The error bars show one standard deviation. The solid lines are the theoretical predictions. \label{fig:comparison}}
\end{figure}

The intensities recorded in the experiment, completely determine $\braket{\hat{\sigma}_z^\mathrm{p}}_\mathrm{w}$. Using $\braket{\hat{\sigma}_z^\mathrm{p}}_\mathrm{w}$ the weak values of the path projection operators $\braket{\hat{\Pi}_{z+}^\mathrm{p}}_\mathrm{w}$ and $\braket{\hat{\Pi}_{z-}^\mathrm{p}}_\mathrm{w}$ are calculated. They in turn are directly related to the preselected path state, making $\Ket{\mathrm{P}_\mathrm{i}^\mathrm{m}}$ available. The results of such a direct state characterization are shown in FIG.~\ref{fig:comparison}, where the weak and the strong interaction approach are compared to each other. The finite contrast of the interferometer has been taken into account for the state characterization.
The upper panels show the normalization factor $\nu$, which connects the probability amplitudes $c_+$ and $c_-$ to $\braket{\hat{\Pi}_{z\pm}^\mathrm{p}}_\mathrm{w}$ for both the weak and the strong interaction case. The middle panels depict the parameter describing the relative weighting $\theta$ of the general preselected state described by Eq.~(\ref{eqn:preselected_general}) again for both measurement approaches. Finally the relative phase $\phi$ is depicted in the lower panels. The lines represent the theoretical prediction. In the results of the state characterization the advantages of the strong interaction approach are evident.

While both measurements are in good agreement with the theoretical predictions, the strong measurement results are significantly better, in terms of precision $\bar{\sigma}$ (a measure of fluctuation) and accuracy $\bar{\Delta}$ (a measure of deviation from the theoretical prediction).\\
For the evaluation of precision of weak and strong interaction approach we use the root mean square statistical error $\bar{\sigma}=\sqrt{\nicefrac{1}{N}\sum_i^N \left|\sigma_i \right|^2}$. The root mean square deviation ${\bar{\Delta}=\sqrt{\nicefrac{1}{N}\sum_i^N \left|t_i-m_i\right|^2}}$ of each measured point $m_i$ from the theoretical predictions $t_i$ in turn is a measure of accuracy. In TABLE~\ref{tab:quant_results} the results of this quantitative analysis are listed.

\begin{table}[h!]
\caption{\label{tab:quant_results}Quantitative comparison of precision $\bar{\sigma}$ and accuracy $\bar{\Delta}$ of the weak and the strong interaction approach.}
\begin{ruledtabular}
\begin{tabular}{ ccc || ccc }
\multicolumn{3}{c ||}{Precision $\bar{\sigma}$}&\multicolumn{3}{c}{Accuracy $\bar{\Delta}$}\\\hline
& Weak  & Strong &  & Weak & Strong  \\\hline
$\nu$ & 0.100 & 0.036 & $\nu$  & 0.152 & 0.062\\\hline
$\theta$ &0.191 & 0.065 &$\theta$ & 0.100 &0.067 \\\hline
$\phi$ & 0.355 & 0.159 & $\phi$ & 0.860 & 0.580 \\
\end{tabular}
\end{ruledtabular}
\end{table}

The strong interaction scheme surpasses the weak one in both accuracy and precision for each and every one of the experimentally determined parameters. For all measured quantities $\bar{\sigma}$ is roughly twice as large in the weak interaction case. Also the mean deviation from the theoretical predictions is smaller for the strong interaction approach. There is another important experimental factor that has to be taken into account: the measurement time. To resolve the small phase shifts between  $\text{I}_{z+}$ and $\text{I}_{z-}$ as well as to discern $\text{I}_{x-}$ from the background long counting times were necessary for the weak interaction. For each point on the weak interaction curve a counting time of 540 seconds was necessary, while 290 seconds were sufficient for the strong one.

Our weak value determination protocol makes it possible to obtain weak values of a two-level quantum system with high accuracy and arbitrary measurement strengths. Increasing the measurement strength in our scheme provides a clear discrimination of small signals from the background. This is in particular significant whenever dealing with low intensities. Our measurement scheme is not limited to the neutron spin and path DOF, but is in fact completely general and can be used for any coupling between two two-level quantum systems. Furthermore, as long as the meter system is two dimensional, it can be used to determine projection operator's weak values of any discrete n-dimensional quantum system.\\
In summary, we have presented a weak value determination scheme via arbitrary interaction strengths. We have applied it to experimentally determine weak values using both weak and strong interactions. We have directly characterized the preselected state of the investigated quantum system including its normalization factor $\nu$, its relative phase $\phi$ and the weight of its eigenstates $\theta$. Experimental evidence is given that strong interactions are superior in terms of accuracy and precision, as well as required measurement time. Our results are not limited to neutron interferometry, but can be used in many n-dimensional quantum systems.

\begin{acknowledgments}
We acknowledge support by the Austrian Science Fund (FWF) Project Nos. P24973-N20 \& P25795-N20. In addition to that we thank the ILL for its hospitality and continuous support. Helpful discussions with Prof. M.~Suda from the Austrian Institute of Technology are kindly acknowledged.
\end{acknowledgments}



\begin{thebibliography}{1}
\bibitem{aav88}  Y.~Aharonov, D.~Z.~Albert, and L.~Vaidman, Phys.~Rev.~Lett. \textbf{60,} 1351 (1988).
\bibitem{duck89} I.~M.~Duck, P.~M.~Stevenson, and E.~C.~G.~Sudarshan, Phys.~Rev.~D \textbf{40,} 2112 (1989).
\bibitem{legget89} A.~J.~Leggett, Phys.~Rev.~Lett. \textbf{62,} 2325 (1989).
\bibitem{svensson2013}B.~E.~Y.~Svensson, Quanta \textbf{2,} 18 (2013).
\bibitem{ferrie2014}C.~Ferrie and J.~Combes, Phys.~Rev.~Lett. \textbf{113,} 120404 (2014).
\bibitem{sokolovski2015}D.~Sokolovski, Phys.~Lett.~A \textbf{379,} 1097 (2015).
\bibitem{pusey2014}M.~F.~Pusey, Phys.~Rev.~Lett. \textbf{113,} 200401 (2014).
\bibitem{dressel2015} J.~Dressel, Phys.~Rev.~A \textbf{91,} 032116 (2015).
\bibitem{dressel2014} J.~Dressel, M.~Malik, F.~M.~Miatto, A.~N.~Jordan, and R.~W.~Boyd, Rev.~Mod.~Phys. \textbf{86,} 307 (2014).
\bibitem{hosten08} O.~Hosten and P.~Kwiat, Science \textbf{319,} 787 (2008).
\bibitem{johansen2004}L.~M.~Johansen, Phys.~Lett.~A \textbf{322,} 298 (2004).
\bibitem{hardy1992} L.~Hardy, Phys.~Rev.~Lett. \textbf{68,} 2981 (1992).
\bibitem{lundeen09} J.~S.~Lundeen and A.~M.~Steinberg, Phys.~Rev.~Lett. \textbf{102,} 020404 (2009).
\bibitem{yokota2009}K.~Yokota, T~.Yamamoto, M.~Koashi, and N.~Imoto, New~J.~Phys. \textbf{11,} 033011 (2009).
\bibitem{aharonov1991}Y.~Aharonov and L.~Vaidman, J.~Phys.~A:~Math.~Gen. \textbf{24,} 2315 (1991).
\bibitem{resch2004}K.~J.~Resch, J.~S.~Lundeen, and A.~M.~Steinberg, Phys.~Lett.~A \textbf{324} 125 (2004).
\bibitem{aharonov13} Y.~Aharonov, S.~Popescu, D.~Rohrlich, and P.~Skrzypczyk, New~J.~Phys. \textbf{15,} 113018 (2013).
\bibitem{denkmayr2014}  T.~Denkmayr, H.~Geppert, S.~Sponar, H.~Lemmel, A.~Matzkin, J.~Tollaksen, and Y.~Hasegawa, Nat.~Commun. \textbf{5,} 4492 (2014).
\bibitem{lundeen2011} J.~S.~Lundeen, B.~Sutherland, A.~Patel, C.~Stewart, and C.~Bamber, Nature \textbf{474,} 188 (2011).
\bibitem{salvail2013} J.~Z.~Salvail, M.~Agnew, A.~S.~Johnson, E.~Bolduc, J.~Leach, and R~.W.~Boyd, Nat.~Photon. \textbf{7,} 316 (2013).
\bibitem{vallone2016}G.~Vallone and D.~Dequal, Phys.~Rev.~Lett. \textbf{116,} 040502 (2016).
\bibitem{zhang2016}Y.-X.~Zhang, S.~Wu, and Z.-B.~Chen, Phys.~Rev.~A \textbf{93,} 032128 (2016).
\bibitem{rauch00} H.~Rauch and S.~A.~Werner, \textit{Neutron Interferometry} (Clarendon, Oxford, 2000).
\bibitem{cronin2009}A.~D.~Cronin, J.~Schmiedmayer, and D.~E.~Pritchard, Rev.~Mod.~Phys. \textbf{81,} 1051 (2009).
\bibitem{arndt2014}M.~Arndt, A.~Ekers, W.~von~Klitzing, and H.~Ulbricht, New~J.~Phys. \textbf{14,} 125006 (2012).
\bibitem{ritchie91}N.~W.~M.~Ritchie, J.~G.~Story, and R.~G.~Hulet, Phys.~Rev.~Lett. \textbf{66,} 1107 (1991).
\bibitem{geppert2014}H.~Geppert, T.~Denkmayr, S.~Sponar, H.~Lemmel, and Y.~Hasegawa, Nucl.~Instr.~Meth.~A \textbf{763,} 417 (2014).
\bibitem{sponar2015}S.~Sponar, T.~Denkmayr, H.~Geppert, H.~Lemmel, A.~Matzkin, J.~Tollaksen, and Y.~Hasegawa, Phys.~Rev.~A \textbf{92,} 062121 (2015).
\bibitem{sponar2016}S.~Sponar, T.~Denkmayr, H.~Geppert, and Y.~Hasegawa, Atoms \textbf{4,} 11 (2016).
\bibitem{rauch02}H.~Rauch, H.~Lemmel, M.~Baron, and R.~Loidl, Nature \textbf{417,} 630 (2002).
\bibitem{hasegawa03}Y.~Hasegawa, R.~Loidl,G.~Badurek, M.~Baron, and H.~Rauch, Nature \textbf{425,} 45 (2003).
\bibitem{pushin2011}D.~A.~Pushin, M.~G.~Huber, M.~Arif, and D.~G.~Cory, Phys.~Rev.~Lett. \textbf{107,} 150401 (2011).
\bibitem{hasegawa11}Y.~Hasegawa and H.~Rauch, New~J.~Phys. \textbf{13,} 115010 (2011).
\bibitem{klepp2014}J.~Klepp, S.~Sponar, and Y.~Hasegawa, Prog.~Theor.~Exp.~Phys. \textbf{2014,} 082A01 (2014).
\bibitem{clark2015}C.~W.~Clark, R.~Barankov, M.~G.~Huber, M.~Arif, D.~G.~Cory, and D.~A.~Pushin, Nature \textbf{525,} 504 (2015).
\bibitem{basu2001}S.~Basu, S.~Bandyopadhyay, G.~Kar, and D.~Home, Phys.~Lett.~A \textbf{279,} 281 (2001).
\bibitem{kroupa2000}G.~Kroupa, G.~Bruckner, O.~Bolik, M.~Zawisky, M.~Hainbuchner, G.~Badurek, R.~J.~Buchelt, A.~Schricker, and H.~Rauch, Nucl.~Instr.~Meth.~A \textbf{440,} 604 (2000).
\end{thebibliography}
\end{document}